\documentclass[prd,preprint,amsmath,nofootinbib,superscriptaddress]{revtex4}
\newcommand{\nn}{\nonumber}

\newcommand{\beq}{\begin{equation}}
\newcommand{\eeq}{\end{equation}}
\newcommand{\bea}{\begin{eqnarray}}
\newcommand{\eea}{\end{eqnarray}}

\begin{document}
\def\pslash{p\!\!\!\slash}

\def\kslash{k\!\!\!\slash}
\def\aslash{a\!\!\!\slash}
\def\bslash{b\!\!\!\slash}
\def\cslash{c\!\!\!\slash}
\def\dslash{\partial\!\!\!\slash}


\title{\boldmath 
Superembedding methods for 4d ${\cal N}=1$ SCFTs\\ }

\author{Walter D. Goldberger}
\affiliation{Department of Physics, Yale University, New Haven, CT 06520}
\affiliation{Physics Department, Columbia University, New York, NY 10027}
\author{Witold Skiba}
\author{Minho Son}
\affiliation{Department of Physics, Yale University, New Haven, CT 06520}

\begin{abstract}
We extend the $SO(4,2)$  covariant lightcone embedding methods of four-dimensional CFTs to ${\cal N}=1$ superconformal field theory (SCFT).   Manifest superconformal $SU(2,2|1)$ invariance is achieved by realizing 4D superconformal space as a surface embedded in the projective superspace spanned by certain complex chiral supermatrices.  Because $SU(2,2|1)$ acts linearly on the ambient space, the constraints on correlators implied by superconformal Ward identities are automatically solved in this formalism.  Applications include new, compact expressions for correlation functions containing one anti-chiral superfield and arbitrary chiral superfield insertions, and manifestly invariant expressions for the  superconformal cross-ratios that parametrize the four-point function of two chiral and two anti-chiral fields.  Superconformal expressions for the leading singularities in the OPE of chiral and anti-chiral operators are also given.  Because of covariance, our expressions should hold in any superconformally flat background, e.g., AdS${}_4$ or ${\bf R}\times {\bf S}^3$.   

 \end{abstract}

\maketitle

\section{Introduction}

The dynamics of four-dimensional conformal field theory (CFT) is tightly constrained by its fifteen parameter group $SO(4,2)$ of spacetime symmetries.    These symmetries completely fix the form of the two-point and three-point correlators of the theory, and reduce the four-point correlators to a function of two conformally invariant cross ratios.    Additional constraints on correlators follow from crossing symmetry, the operator product expansion (OPE), and unitarity, which sets lower bounds on scaling dimensions of fields.

In light of the possible applications of conformal symmetry to strongly coupled models of TeV scale physics and to condensed matter systems, it becomes relevant to fully understand its implications.   A central question~\cite{Polyakov:1974gs} is to what extent does the symmetry algebra, together with general properties such as crossing and unitarity, determine the space of allowed theories, parametrized by the spectrum of primary operators and their OPE coefficients.   

While it is not yet known if such general principles are sufficient to fix the dynamics of CFTs in spacetime dimensions greater than two, there has been recent work that employs  four-dimensional conformal invariance to place non-trivial constraints on the physics.    One such example was put forward in~\cite{Rattazzi:2008pe}, which uses the conformal block structure of four-point functions to derive a sum rule that can be used to place \emph{upper} bounds on the scaling dimensions of certain operators in the theory (see~\cite{Rychkov:2009ij,Poland:2010wg,Rattazzi:2010gj} for extensions of this approach).    These bounds have consequences for model building, as they constrain the types of CFTs that are consistent with the stability of the electroweak scale against large ultraviolet radiative corrections.

Ref.~\cite{Heemskerk:2009pn} provides another recent example of this type of analysis.   This work considers large $N$ four-dimensional CFTs whose spectrum of primary operators contains a relatively  small number of fields with scaling dimensions close to the unitarity bounds (e.g.~conserved currents).    The remaining primaries have dimensions that are parametrically large, so that there is a gap in the spectrum of scaling dimensions (these types of CFTs have been further analyzed by~\cite{Fitzpatrick:2010zm}, which dubs them ``effective CFTs").  As shown in ref.~\cite{Heemskerk:2009pn}, it follows from conformal symmetry that the space of CFTs of this type is in one-to-one correspondence with weakly coupled theories propagating in a background anti-deSitter (AdS) space of one higher dimension.    This result therefore provides a sort of converse to AdS/CFT, and begins to address the question of which type of CFTs can have (even without supersymmetry) weakly coupled gravity duals.

The results uncovered in \cite{Rattazzi:2008pe,Heemskerk:2009pn} make heavy use of the technology of four-dimensional conformal symmetry, in particular the conformal block decomposition of four-point correlators.   If it is indeed possible to sharpen the constraints imposed by $SO(4,2)$ invariance on quantum field theories, it is necessary to develop more efficient methods for representing the observables of a CFT.   A useful language for CFT correlation functions is provided by embedding four-dimensional Minkowski spacetime into the projective lightcone of a six-dimensional flat spacetime with signature $(4,2)$ metric.   The group $SO(4,2)$ acts linearly on the ambient space coordinates, and thus CFT correlators expressed in this language automatically exhibit manifest conformal symmetry.  In addition, results written in terms of the embedding space coordinates are valid not just in Minkowski spacetime, but in any other curved but conformally flat space, as different slicings through the projective lightcone correspond to  conformal transformations of flat space.  This approach goes as far back as ref.~\cite{Dirac:1936fq} for free fields, with applications to more general CFTs given in~\cite{Mack:1969rr,Weinberg:2010fx}.     In the mathematical literature, there exsits a very similar approach to conformal transformations, known as tractor calculus, see e.g~\cite{tractor}.

 The embedding methods have more recently been used  to obtain novel results for CFTs.  These include a simple closed-form expression for the conformal block expansion of four-point functions in spacetimes of dimension $d>2$~\cite{Dolan:2003hv}  (streamlining an earlier approach by the same authors~\cite{Dolan:2000ut}), the formulation of correlators and conformal blocks~\cite{Costa:2011mg} for symmetric traceless primary operators, applications to graviton correlators in conformally flat backgrounds~\cite{Maldacena:2011nz}, and to the Mellin representation of AdS/CFT amplitudes~\cite{Fitzpatrick:2011ia}.

In this paper we extend the six-dimensional embedding space methods to describe ${\cal N}=1$  superconformal field theory (SCFT).   Apart from possible applications of SCFTs to TeV supersymmetry model building, our primary motivation for focusing on such theories is simply that most of the  known four-dimensional interacting exactly conformal theories are also supersymmetric.   Even if there are four-dimensional non-supersymmetric exactly conformal theories, the program of~\cite{Polyakov:1974gs} seems more likely to succeed in the supersymmetric case, where symmetry constraints are stronger.    

After reviewing the embedding formalism for $SO(4,2)$ invariant field theories in Sec.~\ref{sec:rev}, setting up notation for the ${\cal N}=1$ superconformal group $SU(2,2|1)$ in Sec.~\ref{sec:su221} and some of its linear representations in Sec.~\ref{sec:rep}, we turn to the construction of an embedding superspace in Sec.~\ref{sec:6Dsup}.   Our superspace is constructed in such a way that $SU(2,2|1)$ transformations act linearly on the coordinates.  This necessitates the introduction of complex superspace coordinates $(X_{AB}, {\bar X}^{AB})$ constructed in terms of tensor products containing the $SU(2,2|1)$ fundamental $V_A$ and anti-fundamental ${\bar V}^A$.   In Sec.~\ref{sec:4D} we introduce the $SU(2,2|1)$ invariant surface that generalizes the analogous $SO(4,2)$ invariant structure, and show that points on this surface span the standard ${\cal N}=1$ superspace $(x^\mu,\theta,{\bar\theta})$ of four-dimensional supersymmetric field theory.   In Sec.~\ref{sec:fields} we focus our attention on projectively defined holomorphic fields $\Phi(X_{AB})$ and show that they correspond to four-dimensional ${\cal N}=1$ chiral superfields whose $\theta={\bar\theta}=0$ component is an ${\cal N}=1$ SCFT chiral primary operator (an embedding formalism for ${\cal N}=1$ superconformal multiplets with manifest invariance under $SO(4,2)$ but not $SU(2,2|1)$ has been previously introduced in ref.~\cite{Ferrara:1974qk}).    A supersymmetric version of the tractor calculus can be found in~\cite{supertractor}, and a related superembedding formalism for free fields has appeared in ref.~\cite{Siegel:1994cc}.

Some of the consequences for correlators of chiral and anti-chiral supermultiplets are given in Sec.~\ref{sec:apps}.  There, we give manifestly $SU(2,2|1)$ covariant expressions for two-point,  three-point, and in some cases (with only one anti-chiral insertion) $N$-point correlators that are completely fixed by superconformal invariance.  We also introduce an $SU(2,2|1)$ invariant formulation of the OPE of chirals and anti-chirals, and give explicit formulas for the leading OPE singularities (the ${\cal N}=1$ superconformal invariant OPE of currents has been developed, using four-dimensional language, in~\cite{Fortin:2011nq}).   In particular, we recover the additive property of dimensions of operators in the chiral ring.  Finally, we discuss the four-point function with two chiral and two anti-chiral insertions, including its general parameterization in terms of independent coordinate superconformal invariants, and asymptotic behavior in various OPE limits.

In Sec.~\ref{sec:conc} we outline possible extensions of the framework presented here.   In particular, many of the results in this paper should have natural generalizations to extended superconformal symmetry, to supersymmetry in AdS${}_5$, as well to the study of SCFTs in curved,  conformally flat background spacetimes.

\section{Preliminaries}

\subsection{Lightcone methods for CFT}
\label{sec:rev}

The conformal group $SO(4,2)$ realized on four-dimensional Minkowski space is generated by Poincare transformations $(M^{\mu\nu}, P^\mu)$, special conformal transformations $K^\mu$ and dilatations $D$.    While Lorentz transformations and dilatations act linearly on the coordinates $x^\mu$, translations $P^\mu$ and special conformal transformations do not.    For instance, a  finite special conformal transformation parametrized by a four-vector $b^\mu$ sends $x^\mu$ to the point 
\beq
x^\mu \rightarrow \frac{x^\mu + x^2 b^\mu}{1 + 2 b\cdot x + b^2 x^2}.
\eeq

Because conformal transformations act non-linearly on the Poincare coordinates $x^\mu$, the consequences of $SO(4,2)$ invariance for CFT correlation functions can sometimes be obscure in this representation.  This motivates~\cite{Dirac:1936fq,Mack:1969rr,Weinberg:2010fx} the introduction of a set of auxiliary coordinates, which include the Minkowski coordinates $x^\mu$, on which $SO(4,2)$ acts linearly.   Specifically, one introduces a six-dimensional flat spacetime of signature $(4,2)$ with coordinates $X^m$, $m=+,\mu,-$.    The conformal group $SO(4,2)$ acts linearly on these coordinates,
\beq
X^m\rightarrow {\Lambda^m}_n X^n,
\eeq
with $\eta_{mn}  {\Lambda^m}_p  {\Lambda^n}_q =\eta_{pq}$, where the metric is given by
\beq
ds^2 = \eta_{mn} dX^m dX^n = \eta_{\mu\nu} dX^\mu dX^\nu + dX^+ dX^-.
\eeq
For infinitesimal transformations ${\Lambda^m}_n = {\delta^m}_n +{\omega^m}_n$, we have $\delta X^m = \omega^{mn} X_n\equiv  \frac{i}{2} \omega^{pq} L_{pq} X^m$, where the differential operators
\beq
L^{mn}= iX^m \frac{\partial}{\partial X_n} - i X^n  \frac{\partial}{\partial  X_m}
\eeq
generate the $SO(4,2)$ algebra.  Four-dimensional conformally flat spacetime is then recovered as the set of points on the lightcone 
\beq
X^2=\eta_{mn} X^m X^n =0
\eeq
subject to identification of points under re-scalings, $X^m\sim \lambda X^m$ for arbitrary real parameter $\lambda$.  

To see how four-dimensional Minkowski space is embedded in the projective lightcone, consider a slice through the surface $X^2=0$ defined by constant $X^+$.   Introducing the coordinate $x^\mu= X^\mu/X^+,$  we have $X^- = - X^+ x^2$ on the lightcone.    The set of points spanned by all possible coordinate values of $x^\mu$ is then Minkowski space, as can be verified by computing the $SO(4,2)$ transformations of $x^\mu$ induced by those of $X^m$.     First, note that $L^{\mu\nu}$ generates a standard $SO(3,1)$ Lorentz transformation with parameter $\omega^{\mu\nu}$ while
\beq
-iL^{\mu+} \cdot x^\nu = - \eta^{\mu\nu}
\eeq
generates the Poincare translation $\delta x^\mu = \frac{1}{2} \omega^{\mu-}$.   Similarly, $L^{\mu-}$ generates special conformal translations,
\beq
-iL^{\mu-}\cdot x^\nu = \eta^{\mu\nu} x^2 - 2 x^\mu x^\nu,
\eeq
so that $b^\mu = -\omega^{\mu+}/2$.   Finally,
\beq
-iL^{+-}\cdot x^\mu = -2 x^\mu
\eeq
generates dilatations.   Note that the slice with constant $X^+$ defines a good set of coordinates near the ``origin" $X^m=(X^+,X^\mu=0,X^-=0)$.   To describe points near conformal infinity $x^\mu\rightarrow\infty$, it is more suitable to look at slices of constant $X^-$, defining four-dimensional coordinates $z^\mu = X^\mu/X^-$.    These new coordinates are related to $x^\mu$ by the inversion $z^\mu = -x^\mu/x^2$.    It is clear that, given two points on the projective lightcone $X^2=0$, there exists an $SO(4,2)$ transformation that sends one to the origin $x^\mu=0$ and the other to infinity $z^\mu=0$.

It is possible to uplift the action of $SO(4,2)$ not just on the coordinates,  but also on the operator algebra that defines the CFT.   For instance, consider an $SO(4,2)$ scalar operator $\Phi(X)$ defined on the projective lightcone.   Under $SO(4,2)$ transformations,
\beq
\label{eq:6dscalar}
\Phi'(X'=\Lambda \cdot X) = \Phi(X) 
\eeq
and in order for $\Phi(X)$ to be well defined projectively, it must be a homogeneous function of the coordinates
\beq
\label{eq:CFTprim}
\Phi(\lambda X) = \lambda^{-\Delta} \Phi(X),
\eeq
for some real parameter $\Delta$.    Performing an $SO(4,2)$ rotation in the $+-$ plane with $\omega^{+-}=-2\ln\lambda$, we have $X^m=(X^+,X^\mu,X^-)\rightarrow X'{}^m=(\lambda^{-1} X^+, X^\mu, \lambda  X^-)$.    By Eqs.~(\ref{eq:6dscalar}) and~(\ref{eq:CFTprim}),  this implies that $\Phi'(X^+,\lambda  X^\mu, \lambda^{2} X^-) = \lambda^{-\Delta} \Phi(X)$.  Therefore, the four-dimensional operator
\beq
\label{eq:4dCFTprim}
\phi(x^\mu) = (X^+)^\Delta \Phi(X),
\eeq
transforms under such transformations as
\beq
\phi'(\lambda x^\mu) = \lambda^{-\Delta} \phi(x^\mu),
\eeq
so that it corresponds to a CFT operator of definite scaling dimension $\Delta$.   It is also straightforward to check, by applying an $SO(4,2)$ transformation with only $\omega^{\mu+}\neq 0$, that $L^{\mu-} \cdot \phi(0) =0$.  It follows that $\phi(x)$ defined by Eq.~(\ref{eq:4dCFTprim}) corresponds to a CFT primary field of scaling dimension $\Delta$.   This construction has been generalized ~\cite{Mack:1969rr,Weinberg:2010fx} to fields in non-trivial Lorentz representations.  One concludes that the spectrum of primary operators of the CFT is in one-to-one correspondence with fields defined on the projective lightcone.   

One advantage of the lightcone formalism is that conformal invariants and covariants of several points have a simple expression in the six-dimensional embedding space.  The invariants constructed from points $X_{i=1\ldots,N}$ are generated by the scalar products  $X_i\cdot X_j=\eta_{mn} X^m_i X^n_j$ (we will sometimes use the notation $1\cdot 2=X_1\cdot X_2$).   Due to the scaling properties in Eq.~(\ref{eq:CFTprim}), the $N$-point correlator
\beq
\langle \Phi_1(X_1)\cdots \Phi_N(X_N) \rangle
\eeq
can only depend on 
\beq
\left(\begin{array}{c} N\\ 2\end{array}\right)-N = \frac{N (N-3)}{2}
\eeq
independent invariant cross ratios, e.g.
\beq
{(1\cdot 2) (3 \cdot 4)\over (1\cdot 4) (3\cdot 2)} =  {(x_1-x_2)^2 (x_3-x_4)^2\over (x_1-x_4)^2 (x_3-x_2)^2},
\eeq
etc.\ for $N\geq 4$.  (We have used $1\cdot 2 =-{1\over 2} X_1^+ X_2^+ (x_1-x_2)^2$ for points on the lightcone.)  For $N=2$, the two-point function $\langle \Phi_1(X_1) \Phi_2(X_2)\rangle$ is a function of the invariant $X_1\cdot X_2$.  The functional form is fixed by applying a scale transformation $X_1\rightarrow \lambda X_1$ which gives 
\beq
\langle \Phi_1(X_1) \Phi_2(X_2)\rangle = c_{12} {1\over (X_1\cdot X_2)^{\Delta_1}}.
\eeq
On the other hand, using Eq.~(\ref{eq:CFTprim}) with respect to $\Phi_2$, implies that this equation is only consistent if $c_{12}=0$ whenever $\Delta_1\neq\Delta_2$ (in unitary CFTs $c_{12}\geq 0$).  For $\Delta_1=\Delta_2$, the coefficient depends on the normalization of the operators and can be set to unity by a suitable choice of operator basis.   Finally for $N=3$, similar arguments constrain the correlator to have the form
\beq
\langle\Phi_1(X_1) \Phi_2(X_2) \Phi_3(X_3)\rangle = c_{123} (X_2\cdot X_3)^{\Delta_1-\Delta_2-\Delta_3\over 2}   (X_1\cdot X_3)^{\Delta_2-\Delta_1-\Delta_3\over 2} (X_1\cdot X_2)^{\Delta_3-\Delta_1-\Delta_2\over 2}.
\eeq
It follows from this result together with the form of the two-point function (normalized to $c_{12}=1$) that the OPE contains a term 
\beq
\Phi_1(X_1\rightarrow X_2) \cdot \Phi_2(X_2) \sim c_{123}(X_1\cdot X_2)^{\Delta_3-\Delta_1-\Delta_2\over 2} \Phi_3(X_2) + \cdots.
\eeq
Thus, we see how the basic properties of CFTs are recovered in the lightcone embedding approach.

In what follows we formulate the lightcone construction in $SU(2,2)$ rather than $SO(4,2)$ language because it will be useful later on.  The vector $X^m$ of $SO(4,2)$ is equivalent to an antisymmetric tensor of $SU(2,2)$.   If $SU(2,2)$ acts on a defining four-component spinor $V_{\alpha=1,\ldots,4}$ (our spinor conventions are summarized in appendix~\ref{app:spin}) by 
\beq
V_\alpha \rightarrow {U_\alpha}^\beta V_\beta,
\eeq
one obtains the vector of $SO(4,2)$ as the antisymmetric $SU(2,2)$ tensor $X_{\alpha\beta}=-X_{\beta\alpha}$, with each index transforming as in the above equation.    An explicit map between $X^m$ in $SO(4,2)$ and $X_{\alpha\beta}$ transforming as an $SU(2,2)$ tensor can be constructed by introducing a set of matrices $({\Gamma^m})^{\alpha\beta}$ and 
\beq
 {\tilde{\Gamma}^m}_{\alpha\beta} = {1\over 2} \epsilon_{\alpha\beta\rho\sigma} ({\Gamma^m})^{\rho\sigma},
\eeq
where $\epsilon_{\alpha\beta\rho\sigma}$ is the invariant epsilon tensor of $SU(2,2)$.   Then the correspondence is 
\bea
X^m \leftrightarrow {1\over 2} X_{\alpha\beta} {\Gamma^m}^{\alpha\beta}, &  X_{\alpha\beta}  \leftrightarrow \displaystyle{1\over 2} X_m {\tilde{\Gamma}^m}_{\alpha\beta}.
\eea
Useful properties of the matrices ${\Gamma^m}^{\alpha\beta}$ are collected in appendix~\ref{app:spin}.  In $SU(2,2)$ language, the projective lightcone is the surface
\beq
X^2= X_{\alpha\beta} X^{\alpha\beta}=0,
\eeq
with $X^{\alpha\beta} = {1\over 2} \epsilon^{\alpha\beta\rho\sigma} X_{\rho\sigma}$.   The $SU(2,2)$ generators acting on the coordinates are 
\beq
{L_\alpha}^\beta = X_{\alpha\lambda}{\partial\over\partial X_{\beta\lambda}} - {1\over 4} \delta_\alpha{}^\beta X_{\rho\sigma}{\partial\over\partial X_{\rho\sigma}} 
\eeq
and can be related to the operator $L^{mn}$ defined above through
\bea
{L_\alpha}^\beta = - {1\over 2} {(\Sigma^{mn})_\alpha}^\beta L_{mn}, & L^{mn} = -(\Sigma^{mn})_\alpha{}^\beta {L_\beta}^\alpha, 
\eea
where  $(\Sigma^{mn})_\alpha{}^\beta=-{i\over 4}\left({\tilde\Gamma}^m \Gamma^n - {\tilde \Gamma}^n \Gamma^m\right)_\alpha{}^\beta$ are $SO(4,2)$ generators in the spinor representation.

\subsection{The superconformal algebra}
\label{sec:su221}

The ${\cal N}=1$ superconformal group in four spacetime dimensions is the supergroup $SU(2,2|1)$ consisting of supermatrices
\begin{equation}
{U_A}^B =
\left(\begin{array}{cc}
{U_\alpha}^\beta & \phi_\alpha\\
\psi^\beta & z
\end{array}
\right),
\end{equation}
with commuting $4\times4$ and $1\times 1$ blocks  ${U_\alpha}^\beta$ and $z$, respectively, and anticommuting entries $\phi_\alpha,$ $\psi^\beta$.  An object in the fundamental of $SU(2,2|1)$ is given by 
\begin{equation}
\label{eq:defrep}
{V_A} =
\left(\begin{array}{c}
V_\alpha\\
\psi
\end{array}
\right),
\end{equation}
and transforms as $V_A\rightarrow {U_A}^B V_B$ under superconformal transformations.   In order to construct a superspace we will assume later on that $V_\alpha$ is fermionic (anticommuting) and $\psi$ is bosonic.   To formulate the condition that ${U_A}^B$ is an element of $SU(2,2|1)$, note that the bosonic subgroup $SU(2,2)$ is defined to leave invariant the bilinear combination
\begin{equation}
{\bar W}^\alpha V_\alpha = W_{\dot{\alpha}} A^{\dot{\alpha}\alpha} V_\alpha,
\end{equation}
where $W_{\dot{\alpha}} = (W_\alpha)^\dagger$ and $A^{\dot{\alpha}\alpha}$ is the $SU(2,2)$ invariant metric (our conventions for $A^{\dot{\alpha}\alpha}$ are given in appendix~\ref{app:spin}).  Thus we introduce the invariant metric for $SU(2,2|1)$
\begin{equation}
A^{\dot{A}B} =
\left(\begin{array}{cc}
A^{\dot{\alpha}\beta} & 0\\
0  & 1
\end{array}
\right),
\end{equation}
and define  $SU(2,2|1)$ by the condition
\begin{equation}
\label{eq:def}
A^{\dot{A}A} = {U_{\dot{B}}}^{\dot{A}}  A^{\dot{B}B} {U_{B}}^{A},
\end{equation}
where ${U_{\dot{B}}}^{\dot{A}} = ({U_{B}}^{A})^\dagger$.   This equation defines $U(2,2|1)$.   Its subgroup $SU(2,2|1)$ consists of supermatrices with unit superdeterminant,
\begin{equation}
[\mbox{sdet} U]^{-1} = {\mbox{det} ({U_\alpha}^\beta - z^{-1} \phi_\alpha \psi^\beta)\over z}=1.
\end{equation}

To construct the superalgebra~\cite{Kac:1977em}, write an infinitesimal transformation in the fundamental representation as
\begin{equation}
{U_A}^B = {\delta_A}^B  + i {T_A}^B,
\end{equation}
with generator ${T_A}^B$ given by 
\begin{equation}
{T_A}^B =
\left(\begin{array}{cc}
{T_\alpha}^\beta & \phi_\alpha\\
\psi^\beta & \phi
\end{array}
\right).
\end{equation}
In terms of ${T_A}^B$, Eq.~(\ref{eq:def}) becomes
\begin{equation}
\label{eq:alg}
 A^{\dot{A}B} {T_B}^A -  A^{\dot{B}A}  {T_{\dot{B}}}^{\dot{A}}  = 0.
\end{equation}
For indices $A,B=\alpha$ this equation just says that ${T_{\alpha}}^\beta$ is a $U(2,2)$ generator, i.e.
\begin{equation}
 A^{\dot{\alpha}\beta} {T_\beta}^\alpha -  A^{\dot{\beta}\alpha}  {T_{\dot{\beta}}}^{\dot{\alpha}}  = 0.
\end{equation}
For mixed indices, we obtain the reality condition
\begin{equation}
\psi^\beta = {\bar\phi}^\beta \equiv  \phi_{\dot{\alpha}} A^{\dot{\alpha}\beta}, 
\end{equation}
while for $A=B=\psi$,  Eq.~(\ref{eq:alg}) imposes $\phi=\phi^\dagger$.   Finally, the condition $\mbox{sdet} U=1$ implies that ${T_A}^B$ has vanishing supertrace:
\begin{equation}
\mbox{str} {T_A}^B  = {T_\alpha}^\alpha -\phi=0.
\end{equation}
Redefining ${T_\alpha}^\beta$ to remove its trace, so that it becomes a generator of $SU(2,2)$, we end up with the expression
\begin{equation}
{T_A}^B =
\left(\begin{array}{cc}
{T_\alpha}^\beta +{1\over 4}{\delta_\alpha}^\beta \phi & \phi_\alpha\\
{\bar\phi}^\beta & \phi
\end{array}
\right)
\end{equation}
for an $SU(2,2|1)$ generator acting on the fundamental representation.   

Defining 
\begin{equation}
\delta V_A  = i\, {T_A}^B V_B,
\end{equation}
we get component transformation rules
\begin{eqnarray}
-i \delta V_\alpha &=& {T_\alpha}^\beta V_\beta +\phi_\alpha \psi + {1\over 4} \phi V_\alpha,\\
-i \delta \psi           &=&  {\bar \phi}^\beta V_\beta + \phi \psi.
\end{eqnarray}
In particular, one can define supersymmetry generators $Q_\alpha$ and ${\bar Q}^\beta$ such that
\begin{eqnarray}
-i \delta \left(\begin{array}{c}
V_\alpha\\
\psi
\end{array}
\right)
=  \left[{\bar\phi}^\beta Q_\beta + \phi_\beta {\bar Q}^\beta\right] \cdot \left(\begin{array}{c}
V_\alpha\\
\psi
\end{array}
\right).
\end{eqnarray}
Thus as supermatrices in the fundamental representation
\begin{eqnarray}
{(Q_\sigma)_A}^B =\left(\begin{array}{cc}
0 &0\\
{\delta_\sigma}^\beta & 0
\end{array}
\right), & 
{({\bar Q}^\sigma)_A}^B =\left(\begin{array}{cc}
0 & {\delta_\alpha}^\sigma\\
0 & 0
\end{array}
\right).
\end{eqnarray}
If in addition we define $U(1)_R$ acting on the fundamental representation by
\begin{equation}
-i \delta \left(\begin{array}{c}
V_\alpha\\
\psi
\end{array}
\right) = \phi R \cdot \left(\begin{array}{c}
V_\alpha\\
\psi
\end{array}
\right),
\end{equation}
with
\begin{equation}
{R_A}^B = \left(\begin{array}{cc}
{1\over 4} {\delta_\alpha}^\beta & 0\\
0& 1
\end{array}
\right),
\end{equation}
or $R(V_\alpha)=1/4$, and $R(\psi)=1$.   The normalization chosen for $U(1)_R$ is an arbitrary convention.  We will relate this to the standard normalization for $R$-charge in Sec.~\ref{sec:fields}.    For $SU(2,2)$ a convenient basis of generators is
\begin{equation}
\label{eq:J}
{[{J_\alpha}^\beta]_\rho}^\sigma = {\delta_\alpha}^\sigma  {\delta_\rho}^\beta - {1\over 4}{\delta_\alpha}^\beta {\delta_\rho}^\sigma.
\end{equation}

Given the explicit matrix representation, the algebra follows.   The commutator of two $J$'s just gives the $SU(2,2)$ algebra,
\beq
[{J_\alpha}^\beta,{J_\rho}^\sigma]=\delta_\alpha{}^\sigma J_\rho{}^\beta - \delta_\rho{}^\beta  J_\alpha{}^\sigma,
\eeq
while the other relations are 
\begin{eqnarray}
\label{eq:r}
[R,{J_\alpha}^\beta]=0, & [R,Q_\alpha] = \displaystyle{{3\over 4}}  Q_\alpha  ,& [R,{\bar Q}^\alpha] =- {3\over 4} {\bar Q}^\alpha,
\end{eqnarray}
and 
\begin{equation}
\label{eq:qq}
\{Q_\alpha, {\bar Q}^\beta\} = {J_\alpha}^\beta + {\delta_\alpha}^\beta R.
\end{equation}
We will relate this manifestly $SU(2,2)$ covariant form of the ${\cal N}=1$ superconformal generators to the more conventional (Poincare covariant) form in Sec.~\ref{sec:4D}.

\subsection{Representations and invariants}
\label{sec:rep}

Starting from the fundamental representation $V_A$, transforming as 
\begin{equation}
V_A\rightarrow {U_A}^B V_B,
\end{equation}
and its conjugate ${\bar V}^A = A^{A{\dot B}} V_{\dot B}$, with transformation law
\begin{equation}
\label{eq:conj}
{\bar V}^A\rightarrow  {\bar V}^B  {(U^{-1})_B}^A,
\end{equation}
we can form several product representation that will be useful in our construction of the superspace below.  A more complete discussion of linear $SU(2,2|1)$ representations can be found in refs.~\cite{Gursey:1977ag,Flato:1983te,Dobrev:1985qv} (see~\cite{Minwalla:1997ka} for a review).   

From the definitions, the product ${Z_A}^B= V_A {\bar W}^B$ of a fundamental $V_A$ and an anti-fundamental ${\bar W}^B$ transforms as a tensor
\begin{equation}
\label{eq:adj}
{Z_A}^B\rightarrow  {U_A}^C {Z_C}^D {(U^{-1})_D}^B.
\end{equation}
This is a reducible representation, containing the singlet
\begin{equation}
{\bar W}^A V_A  = {\bar W}^A  {\delta_A}^B V_B\equiv  V_B {\lambda_A}^B {\bar W}^A= {Z_A}^B  {\lambda_A}^B ,
\end{equation}
where we have introduced
\begin{equation}
{\lambda_A}^B = 
\left(\begin{array}{cc}
-{\delta_\alpha}^\beta & \\
& 1
\end{array}\right).
\end{equation}    
In particular, the supertrace is given by  $\mbox{str}\,  {T_A}^B = - {T_A}^B {\lambda_B}^A$.   The product also contains the adjoint representation, the super-traceless part of ${Z_A}^B$,
\begin{equation}
 {Z_A}^B  +{1\over 3}({Z_C}^D  {\lambda_D}^C) {\delta_A}^B.
\end{equation}

In addition, one can construct tensors $X_{AB}$ and ${\bar X}^{AB}$ which transform in the same way as the products $V_A V_B$ and ${\bar V}^A {\bar V}^B$ respectively.   Their transformation properties under $SU(2,2|1)$ are
\begin{eqnarray}
\label{eq:xtran}
X_{AB} &\rightarrow& {U_A}^C X_{CD} {{\hat U}}_B{}^D,\\
\label{eq:xbtran}
{\bar X}^{AB} &\rightarrow&  ({\hat U^{-1}})_C{}^A {\bar X}^{CD} {(U^{-1})_D}{}^B,
 \end{eqnarray}
 where ${{\hat U}}_A{}^B$ is obtained from ${{U}}_A{}^B$ by setting ${\hat U}_\psi{}^\delta = - U_\psi{}^\delta$ and keeping all other matrix elements the same, while $({\hat U^{-1}})_A{}^B$ is given by  $({\hat U^{-1}})_\alpha{}^\psi = - (U^{-1})_\alpha{}^\psi$.    It then follows that $X_{AC}{\lambda_D}^C {\bar X}^{DB}$ transforms as in Eq.~(\ref{eq:adj}), and due to the property $\mbox{str} M_1 M_2 =\mbox{str} M_2 M_1$, the supertrace
\beq
\mbox{str} X \lambda {\bar Y} = - \lambda_B{}^A X_{AC} \lambda_D{}^C {\bar Y}^{DB}
 \eeq     
 is superconformally invariant.

\section{Superspace}
\label{sec:6Dsup}

Our goal in this section is to construct a superconformal extension of the six-dimensional projective lightcone.  In order to do this, we use the equivalence between the $SO(4,2)$ vector $X^m$ and the $SU(2,2)$ anti-symmetric tensor $X_{\alpha\beta}$.   The latter transforms under $SU(2,2)$ as the product $V_\alpha V_\beta$ of an \emph{anticommuting} $SU(2,2)$ spinor $V_\alpha$.   This suggests that the supersymmetric extension of $X_{\alpha\beta}$ is a tensor $X_{AB}$ transforming as the product 
\beq
X_{AB} \sim V_A V_B,
\eeq
where $V_A$ is the object introduced in Eq.~(\ref{eq:defrep}).   The coordinate supermultiplet $X_{AB}$ contains the usual $SU(2,2)$ coordinate $X_{\alpha\beta}=-X_{\beta\alpha}$, as well as its fermionic superpartner
\beq
\theta_\alpha = X_{\alpha\psi} = X_{\psi\alpha},
\eeq
which transforms as an $SU(2,2)$ spinor. In addition, $X_{AB}$ contains the bosonic coordinate
\beq
\varphi = X_{\psi\psi}.
\eeq
The objects $V_A$,  $X_{AB}$ introduced here are analogous to quantities appearing in the $SU(2,2|3)$ superconformal twistor construction of Ref.~\cite{Sinkovics:2004fm}, see also Ref.~\cite{Kuzenko:2006mv}.   It is not possible to impose an $SU(2,2|1)$ invariant reality condition on $X_{AB}$ and we are forced to introduce the conjugate, denoted by ${\bar X}^{AB}$, as well.   It transforms as the product ${\bar V}^A {\bar V}^B$ of two $SU(2,2|1)$ anti-fundamental representations, introduced in Eq.~(\ref{eq:conj}).    The complete superspace is then the space spanned by the pair $(X_{AB}, {\bar X}^{AB})$.    We impose the identification $(X_{AB}, {\bar X}^{AB})\sim (\lambda X_{AB}, {\bar\lambda}{\bar X}^{AB})$ for arbitrary $\lambda\in{\bf C}$.   The coordinates $X_{AB}$ and ${\bar X}^{AB}$ are independent, although below we will identify ${\bar X}^{AB}$ with the conjugate of $X_{AB}$.  In components, this means
\beq
{\bar X}^{\alpha\beta} = - A^{{\dot\alpha}\alpha} A^{{\dot\beta}\beta} X_{{\dot\alpha}{\dot\beta}},
\eeq
where $X_{{\dot\alpha}{\dot\beta}}= (X_{\alpha\beta})^\dagger$, 
\beq
{\bar\theta}^\alpha = A^{{\dot\alpha}\alpha} \theta_{{\dot\alpha}},
\eeq
and ${\bar\varphi}=\varphi^\dagger$.

The objects $X_{AB}$ and ${\bar X}^{AB}$ transform under $SU(2,2|1)$ according to Eqs.~(\ref{eq:xtran}) and (\ref{eq:xbtran}), respectively.   It follows that there is an $SU(2,2|1)$ invariant inner product between points on our superspace, given by 
\beq
\label{eq:dot}
X_1 \cdot {\bar X}_2 \equiv \mbox{str} (X_1\lambda{\bar X}_2) =  - (X_1)_{AB} \lambda_C{}^B ({\bar X}_2)^{CD} \lambda_D{}^A.
\eeq
This scalar product will play an important role in constructing superconformally covariant correlation functions later on.   In components, it is given by
\beq
X_1 \cdot {\bar X}_2 = X_1{}_{\alpha\beta} {\bar X}_2^{\alpha\beta}  + 2 {\theta_1}{}_\alpha {\bar \theta}_2^{\alpha} - \varphi_1 {\bar\varphi}_2.
\eeq
The existence of a scalar product also implies that our superspace has a natural invariant supermetric  
\beq
ds^2 = dX\cdot d{\bar X}  =   - dX_{AB} \lambda_C{}^B d{\bar X}^{CD} \lambda_D{}^A,
\eeq
whose specific role we will not explore further in this paper.

From the transformation of $V_A$ in the previous section, or from Eqs.~(\ref{eq:xtran}) and~(\ref{eq:xbtran}), it is possible to work out the superconformal transformation of the component coordinates
\begin{eqnarray}
\label{eq:t1st}
-i\delta X_{\alpha\beta} &=&  {T_\alpha}^\sigma X_{\sigma\beta}  + {T_\beta}^\sigma X_{\alpha\sigma} + \phi_\alpha \theta_\beta - \phi_\beta\theta_\alpha+ {1\over 2}\phi X_{\alpha\beta},\\
-i\delta\theta_\alpha &=&  {T_\alpha}^\sigma \theta_\sigma+  {\bar\phi}^\sigma X_{\sigma\alpha} + \phi_\alpha \varphi + {5\over 4} \phi \theta_\alpha,\\
-i\delta\varphi            &=&  2 {\bar\phi}^\beta \theta_\beta + {2} \phi\varphi.
\end{eqnarray}
Likewise,
\begin{eqnarray}
i\delta {\bar X}^{\alpha\beta} &=&  {T_\sigma}^\alpha {\bar X}^{\sigma\beta}  +  {T_\sigma}^\beta {\bar X}^{\alpha\sigma} + {\bar \phi}^\alpha {\bar\theta}^\beta - {\bar \phi}^\beta{\bar\theta}^\alpha + {1\over 2}\phi {\bar X}^{\alpha\beta},\\
i\delta{\bar\theta}^\alpha &=&  {T_\sigma}^\alpha {\bar\theta}^\sigma - {\phi}_\sigma {\bar X}^{\sigma\alpha} + {\bar\phi}^\alpha {\bar\varphi} + {5\over 4} \phi {\bar\theta}^\alpha,\\
\label{eq:tlast}
i\delta{\bar\varphi}            &=&  -2 {\phi}_\sigma {\bar \theta}^\sigma + {2} \phi\bar{\varphi}.
\end{eqnarray}
One can check that these transformations are consistent with the superalgebra in Eqs.~(\ref{eq:r}) and~(\ref{eq:qq}).    From these expressions we can also obtain expressions for the differential operators realizing the algebra on superspace.    These results can be found in appendix~\ref{app:diff}.

\subsection{4D superspace}
\label{sec:4D}

Just as Minkowski spacetime arises as (a portion) of the six-dimensional projective lightcone $X_m X^m =0$, four-dimensional superspace can be described as a superconformally invariant surface embedded in the superspace $(X_{AB}, {\bar X}^{AB})$.   The correct surface should be the supersymmetric extension of an $SU(2,2)\times U(1)_R$ invariant constraint such as $X_{\alpha\beta} X^{\alpha\beta}=0$ or $X_{\alpha\beta} {\bar X}^{\alpha\beta}=0$.    In order to determine the correct constraints we will start with the definition of four-dimensional superspace $(x^\mu,\theta_a,{\bar\theta}^{\dot a})$ as the coset $SU(2,2|1)/H$~\cite{sohnius}, where $H$ consists of the ${\cal N}=1$ super-Poincare group with generators $J^{\mu\nu}$, $K^\mu$, $S_a$,  ${\bar S}^{\dot a}$,  together with $D$ and $R$.    This coset consists of the set of points generated by applying all possible $SU(2,2|1)$ transformations to the ``origin"
\bea
{\hat X}_{\alpha\beta} = {1\over 2}\left(\begin{array}{cc}
i\epsilon_{ab} X^+ & 0\\
0 & 0
\end{array}\right), & 
\theta_\alpha =0 ,
&
\varphi =0,
\eea
and likewise for the complex coordinate ${\bar X}^{AB}$.

Explicitly, note that the isotropy group of the origin  consists of $SU(2,2)$ transformations with $\omega^{\mu-}=0$ (i.e. Lorentz  and special conformal transformations) and by supersymmetry transformations parametrized by spinors of the form
\beq
\phi_\alpha = \left(\begin{array}{c}
\eta_a \\
0
\end{array}\right).
\eeq
Note that the origin is also projectively invariant under dilatations (generated by $\omega^{+-}\neq 0$) and by $U(1)_R$ transformations.    Thus non-trivial solutions are parameterized by translations $\omega^{\mu-} = 2 x^\mu$ and by supersymmetry transformations with parameter
\beq
\phi_\alpha = \left(\begin{array}{c}
0 \\
2 {\bar\theta}^{\dot a}
\end{array}\right).
\eeq
The explicit $SU(2,2|1)$ transformation that generates the point labeled by $(x^\mu,{\theta}_a,{\bar\theta}^{\dot a})$ can be decomposed as
\beq
{U(x,\theta,{\bar\theta})_A}^B = U(x,0,0)_A{}^C U(0,\theta,{\bar \theta})_C{}^B,
\eeq
with 
\beq
U(x,0,0)_A{}^B =\left(
\begin{array}{cc}
\left(e^{-i x_\mu \Sigma^{\mu +}}\right)_\alpha{}^\beta & 0\\
0 																				   & 1
\end{array}\right)
=\left(
\begin{array}{ccc}
\delta_a{}^b & 0 & 0\\
i x^\mu ({\bar\sigma}_\mu)^{{\dot a}  b} & \delta^{\dot a}{}_{\dot b} & 0\\
0														& 0									& 1
\end{array}\right),
\eeq
and
\beq
\label{eq:ttrans}
U(0,\theta,{\bar \theta})_A{}^B =\left(
\begin{array}{cc}
\delta_\alpha{}^\beta -{1\over 2} \phi_\alpha {\bar\phi}^\beta& i \phi_\alpha\\
i{\bar \phi}^\beta																				   & 1
\end{array}\right)
=\left(
\begin{array}{ccc}
\delta_a{}^b & 0 & 0\\
-\theta\sigma^\mu{\bar\theta}  ({\bar\sigma}_\mu)^{{\dot a} b} & \delta^{\dot a}{}_{\dot b} & 2 i {\bar\theta}^{\dot a}\\
2 i \theta^b														& 0									& 1
\end{array}\right).
\eeq
It follows that 
\beq
{U(x,\theta,{\bar\theta})_A}^B = \left(
\begin{array}{ccc}
\delta_a{}^b & 0 & 0\\
i y^\mu  ({\bar\sigma}_\mu)^{{\dot a} b} & \delta^{\dot a}{}_{\dot b} & 2 i {\bar\theta}^{\dot a}\\
2 i \theta^b														& 0									& 1
\end{array}\right),
\eeq
where we have introduced $y^\mu = x^\mu + i \theta \sigma^\mu {\bar\theta}$.  We therefore find
\beq
X_{\alpha\beta}(y,\theta) = U_\alpha{}^\sigma {\hat X}_{\sigma\rho} U_\beta{}^\rho = {1\over 2} Y_m {\tilde\Gamma^m}_{\alpha\beta},
\eeq
with 
\beq
\label{eq:Y}
Y^m = (Y^+ = X^+, Y^\mu = X^+ y^\mu, Y^- = - X^+ y^2),
\eeq
so that $X_{\alpha\beta}$ obeys the ordinary lightcone constraint $X_{\alpha\beta} X^{\alpha\beta}=0$.   In addition,
\beq
\label{eq:theta}
\theta_\alpha(y,\theta) = - U_\alpha{}^\sigma {\hat X}_{\sigma\rho} U_\psi{}^\rho = X^+ \left(\begin{array}{c}
\theta_a\\
i y^\mu  ({\bar\sigma}_\mu)^{{\dot a} b} \theta_b
\end{array}\right)
\eeq
and
\beq
\label{eq:phi}
\varphi(y,\theta) = - U_\psi{}^\alpha {\hat X}_{\alpha\beta} U_\psi{}^\beta = 2 i X^+ \theta^2.   
\eeq
Likewise, we obtain for the complex conjugates, with ${\bar Y}^m = (Y^m)^\dagger$,
\bea
{\bar X}^{\alpha\beta} &=& {1\over 2} {\bar Y}^m {\Gamma_m}^{\alpha\beta},\\
{\bar \theta}^\alpha &=&  {\bar X}^+ \left(\begin{array}{c}
-i {\bar y}^\mu  ({\bar\sigma}_\mu)^{{\dot b} a} {\bar \theta}_{\dot b}\\
{\bar\theta}_{\dot a}
\end{array}\right),\\
{\bar\varphi} &=& - 2i{\bar X}^+ {\bar\theta}^2 .
\eea

The solutions $X_{AB}(y^\mu,\theta)$, ${\bar X}^{AB}({\bar y}^\mu, {\bar\theta})$ written above can be obtained as points obeying a set of algebraic constraints in the embedding superspace.    First of all, $X_{AB}(y^\mu,\theta)$ satisfies the equations
\bea
\label{eq:c1}
X_{\alpha\beta} X^{\alpha\beta} &=&0\\
X^{\alpha\beta} \theta_\beta &=& 0\\
\label{eq:con}
\varphi X_{\alpha\beta} + 2 \theta_\alpha \theta_\beta &=&0\\
\varphi \theta_\alpha &=&0,\\
\label{eq:c5}
\varphi^2 &=&0,
\eea
and similarly for the conjugate coordinates.   These constraints can be generated by applying the differential operators ${\bar Q}^\alpha$ to the lightcone constraint $Z=X_{\alpha\beta} X^{\alpha\beta}=0$.   In fact the constraints fit into a sixteen dimensional irreducible representation of $SU(2,2|1)$ transforming as the product $V_A V_B V_C V_D$.   Thus the constraints satisfied by $X_{AB}$ can be expressed covariantly as $[X_{AB} X_{CD}]_{\bf 16}=0$ and similarly for the conjugates.

It is also possible to start with the constraints in Eqs.~(\ref{eq:c1})-(\ref{eq:c5}) and show that they their most general solution is exactly of the form given in Eqs.~(\ref{eq:Y})-(\ref{eq:phi}), and likewise for the conjugates.    However, this by itself is not sufficient to prove that the coset space $SU(2,2|1)/H$ is given by the equations  $[X_{AB} X_{CD}]_{\bf 16}=[{\bar X}^{AB} {\bar X}^{CD}]_{{\bar\bf 16}}=0$ since in order to send $X_{AB}$ and ${\bar X}^{AB}$ to the origin requires the action of the complexified superconformal group rather than $SU(2,2|1)$ (in particular, sending $X^m ={1\over 2} X_{\alpha\beta} (\Gamma^m)^{\alpha\beta}$ to $X^m=(X^+,0,0)$ requires a complex Poincare translation).   Consequently, additional constraints that relate $X_{AB}$ and ${\bar X}^{AB}$ are needed.

The correct additional constraint turns out to be that $[X_{AB} {\bar X}^{CD}]_{\bf 24}=0$, or explicitly that ${Z_A}^B=0$, where
\begin{equation}
{Z_A}^B = X_{AC}{\lambda_D}^C {\bar X}^{DB} - {1\over 3}{\delta_A}^B \mbox{str} X\lambda {\bar X}
\end{equation}
transforms in the adjoint representation of $SU(2,2|1)$.   Expanding in components, the constraints are
\bea
\label{eq:yybar}
X^m {\bar X}^n - X^n {\bar X}^m &=& 2 i {\bar\theta} \Sigma^{mn} \theta,\\
X_{\alpha\sigma} {\bar\theta}^\sigma - {\bar \varphi} \theta_\alpha &=& 0,\\
{\bar X}^{\alpha\sigma} \theta_\sigma +\varphi {\bar\theta}^\alpha &=& 0,\\
X_{\alpha\beta} {\bar X}^{\alpha\beta} + 5 \theta_\alpha {\bar\theta}^\alpha - 4 \varphi{\bar\varphi} &=&0.
\eea
Inserting the solutions to the holomorphic constraints $[X_{AB} X_{CD}]_{\bf 16}=[{\bar X}^{AB} {\bar X}^{CD}]_{\bar{\bf 16}}=0$ into these equations, the $m=\mu,n=+$ component of Eq.~(\ref{eq:yybar}) implies the relation
\beq
\label{eq:4dchi}
y^\mu - {\bar y}^\mu = 2 i \theta \sigma^\mu {\bar\theta},
\eeq
while all other constraints contained in $[X_{AB} {\bar X}^{CD}]_{\bf 24}=0$ are automatically satisfied once Eq.~(\ref{eq:4dchi}) is imposed.   In addition, points obeying $[X_{AB} X_{CD}]_{\bf 16}=[{\bar X}^{AB} {\bar X}^{CD}]_{\bar{\bf 16}}=0$  and $[X_{AB} {\bar X}^{CD}]_{\bf 24}=0$ automatically obey the singlet  ``superlightcone" constraint $X\cdot{\bar X}=0$.   Finally, identifying  ${\bar X}^{AB}$ with the conjugate of $X_{AB}$ via the $SU(2,2|1)$ covariant condition ${\bar X}^{AB} = A^{{\dot A} A} A^{{\dot B} B} X_{{\dot A} {\dot B}}$ implies that the space of solutions to the ${\bf 16}$, ${\bar{\bf 16}}$ and ${\bf 24}$ constraint is exactly the desired four-dimensional coset superspace\footnote{The need for the constraints $[X_{AB} X_{CD}]_{\bf 16}=[{\bar X}^{AB} {\bar X}^{CD}]_{\bar{\bf 16}}=0$ and $[X_{AB} {\bar X}^{CD}]_{\bf 24}=0$ was first pointed out in~\cite{MS}.   Note that any further quadratic constraints, namely those involving $[X_{AB} X_{CD}]_{\bf 46}$ and $[X_{AB} X_{CD}]_{\bf 59}$ or $[X_{AB} {\bar X}^{CD}]_{\bf 96}$ are too constraining and do not yield the correct four-dimensional space.} parameterized by the coordinates $(x^\mu=(y+{\bar y})^\mu/2,\theta,{\bar\theta})$, see e.g.~\cite{Buchbinder:1998qv}.

Given the transformation properties of $X_{AB}$ worked out above, it is now straightforward to determine how the parameters $(y^\mu,\theta_a)$ transform under superconformal transformations.   The result is that a transformation with parameter
\beq
\label{eq:fp}
\phi_\alpha = \left(\begin{array}{c}
-2 \eta_a \\
2 {\bar\epsilon}^{\dot a}
\end{array}\right),
\eeq
induces the transformation
\bea
\label{eq:4ddeltay}
\delta y^\mu &=& i(\phi_\alpha {\bar Q}^\alpha + {\bar\phi}^\alpha Q_{\alpha}) \cdot {Y^\mu\over Y^+} = 2 i \theta\sigma^\mu {\bar\epsilon} - 2y_\nu \theta \sigma^\mu {\bar\sigma}^\nu \eta,\\
\label{eq:4ddeltatheta}
\delta\theta^a &=& i(\phi_\alpha {\bar Q}^\alpha + {\bar\phi}^\alpha Q_{\alpha}) \cdot \theta^a= \epsilon^a -  i y^\mu ({\bar\sigma}_\mu)^{{\dot b} a} {\bar\eta}_{\dot b} + 2 \theta^2 {\eta}^a,
\eea
which exactly coincides with the supersymmetry transformations of four-dimensional chiral ${\cal N}=1$ superspace $(y^\mu,{\theta}_a)$ (see e.g.~ref.~\cite{Buchbinder:1998qv}).   This establishes the equivalence between points $X_{AB}$ subject to the above covariant constraints and four-dimensional chiral superspace, at least near the origin.    The full superspace, obtained by including the conjugate variable ${\bar X}^{AB}$, is labeled by $(x^\mu = (y+{\bar y})^\mu/2,\theta,{\bar \theta})$. Analogous results have been obtained using the super-twistor formalism, see Refs.~\cite{Kuzenko:2006mv,Siegel:2010yd}.  The relation between our fermionic generator $Q_\alpha$ and the more conventional Poincare supercharge $Q_a$ and special-superconformal supercharge $S_a$ is then
\beq
Q_\alpha = {i\over 2}\left(\begin{array}{c}
-Q_a \\
{\bar S}^{\dot a}
\end{array}\right).
\eeq

We conclude this section by noting that the point $X_{AB}(y,\theta)$ defined by the action of $U_A^B(x,\theta,{\bar\theta})$ on the ``origin" can also be reached starting at ``infinity",
\bea
{\check X}_{\alpha\beta} = {1\over 2}\left(\begin{array}{cc}
0 & 0\\
0 & i\epsilon_{{\dot a}{\dot b}} X^- 
\end{array}\right), & 
\theta_\alpha =0 ,
&
\varphi =0,
\eea
by applying a combination of special conformal and special superconformal transformations.   Such transformations yield a point that is projectively equivalent to $X_{AB}(x,\theta,{\bar\theta})$.   This result indicates in particular that there exists a superconformal transformation that can simultaneously send any two points to the origin and to infinity, respectively.

\section{The chiral sector of SCFTs}
\label{sec:fields}

As in the case of $SO(4,2)$ invariant field theory, the operator content of a four-dimensional SCFT has a covariant description in terms of fields living on the projective ``superlightcone" structure introduced above.    In this section we formulate the correspondence between superfields on the superlightcone and superconformal multiplets in four dimensions.

We start with superfields $\Phi(X,{\bar X})$ satisfying
\beq
\label{eq:scale}
\Phi(\lambda X, {\bar\lambda} {\bar X}) = \lambda^{-\Delta/2} {\bar\lambda}^{-{\bar\Delta}/2} \Phi(X,  {\bar X}),
\eeq
for arbitrary complex parameter $\lambda$.   The superfield $\Phi$ can be assigned to any linear representation of $SU(2,2|1)$, although in this paper we will focus for simplicity on scalars
\beq
\Phi'(X',{\bar X}') = \Phi(X,{\bar X}),
\eeq
with $X'$ and ${\bar X}'$ given by the RHS of Eqs.~(\ref{eq:xtran}) and (\ref{eq:xbtran}).   More general tensors, for instance the important case of the supercurrent multiplet, will be considered elsewhere.   

The scalar superfield $\Phi(X,{\bar X})$ is reducible, as the holomorphy condition $\partial\Phi/\partial{\bar X}=0$ is $SU(2,2|1)$ invariant.   For such a holomorphic field, the condition in Eq.~(\ref{eq:scale}) now becomes simply $\Phi(\lambda X) = \lambda^{-\Delta} \Phi(X)$.    We will now focus on holomorphic fields and show that, upon projection to four dimensions,  they correspond to  SCFT multiplets whose lowest component is a chiral primary operator.   Expanding in powers of $\theta$, $\Phi(X)$ has the component field expansion
\beq
\Phi(X_{AB}) = A + \sqrt{2} \theta_\alpha \psi^\alpha  + {1\over 2!} \theta_\alpha \theta_\beta V^{\alpha\beta} + {1\over 3!} \theta_\alpha \theta_\beta \theta_\sigma \epsilon^{\alpha\beta\rho\sigma} \chi_\sigma+ {1\over 4!} \theta_\alpha \theta_\beta \theta_\sigma \theta_\rho \epsilon^{\alpha\beta\rho\sigma} B,
\eeq
where the component fields are functions of $(X_{\alpha\beta},\varphi)$.   It follows from the scaling law $\Phi(\lambda X_{AB}) = \lambda^{-\Delta} \Phi(X_{AB})$ that the component fields satisfy $A(\lambda X_{\alpha\beta},\lambda \varphi) = \lambda^{-\Delta} A(X_{\alpha\beta}, \varphi)$, $\psi(\lambda X_{\alpha\beta},\lambda \varphi) = \lambda^{-\Delta-1} \psi(X_{\alpha\beta}, \varphi),$ etc.   We may therefore define a four-dimensional superfield, dependent only on the chiral coordinates $(y^\mu,\theta_a)$, by 
\beq
\label{eq:4dchiral}
\Phi(y^\mu,\theta_a) = (X^+)^\Delta \Phi(X_{AB})= A(y) + \sqrt{2}\theta\psi(y) + \theta^2 F(y).
\eeq
Restricted to the four-dimensional superspace, terms with more than two powers of $\theta_\alpha$ vanish, and using $\theta_\alpha \theta_\beta = -i \theta^2 X^+  X_{\alpha\beta}$ from Eq.~(\ref{eq:con}) we obtain the explicit relations 
\bea
A(y) &=& (X^+)^\Delta A(X_{\alpha\beta},0),\\
\psi^a(y) &=& - (X^+)^{\Delta+1}\left[ \psi^{\alpha=a}(X_{\alpha\beta},0) + i y^\mu ({\bar\sigma}_\mu)^{{\dot b} a} \psi_{\alpha={\dot b}}(X_{\alpha\beta},0)\right],
\\
F(y)  &=& i (X^+)^{\Delta+1}\left[2 {\partial\over\partial\varphi} A(X_{\alpha\beta},0) -{1\over 2} X_{\rho\sigma} V^{\rho\sigma}(X_{\alpha\beta},0)\right].
\eea

The transformation of $\Phi(y^\mu,\theta_a)$ under superconformal symmetries is fixed by the definition in Eq.~(\ref{eq:4dchiral}) together with the transformation rule  $\Phi'(X')=\Phi(X)$.   Because $X^+$ is invariant under Poincare supersymmetry, Eqs.~(\ref{eq:4ddeltay}) and~(\ref{eq:4ddeltatheta}) imply that that the component fields $(A(y),\psi(y),F(y))$ transform exactly as a  chiral multiplet of four-dimensional ${\cal N}=1$ Poincare supersymmetry.   To compute the special superconformal transformations of the components, note that $\delta X^+ = - 4 X^+\theta\eta$ for a transformation with parameter as in Eq.~(\ref{eq:fp}).   Then $\delta \Phi(y,\theta) \equiv \Phi'(y,\theta) - \Phi(y,\theta)$ is given by 
\beq
\delta \Phi(y,\theta)=  -4\Delta\theta\eta \Phi(y,\theta) + 2 y_\nu \theta\sigma^\mu {\bar\sigma}^\nu\eta \partial_\mu \Phi(y,\theta)  +  \left(i y^\mu ({\bar\sigma}_\mu)^{{\dot b}a}{\bar\eta}_{\dot b} - 2 \theta^2\eta^a\right){\partial_a} \Phi(y,\theta)
\eeq
which implies that the components transform as
\bea
\delta A(y) &=& i\sqrt{2} y^\mu {\bar\eta} {\bar\sigma}_\mu \psi,\\
\delta \psi_a(y) & =& -2 \sqrt{2} \Delta \eta_a A + \sqrt{2} y_\nu (\sigma^\mu {\bar\sigma}^\nu \eta)_a \partial_\mu A -  \sqrt{2} i y^\mu (\sigma_\mu {\bar \eta})_a F\\
\delta F(y) &=& 2 \sqrt{2} (\Delta- 1) \eta\psi - {\sqrt{2}} y_\nu \partial_\mu\psi \sigma^\mu {\bar\sigma}^\nu \eta.
\eea
In particular we note that at the origin $\delta A(0)=0$, so that $A(0)$ is annihilated by the special superconformal generators $S_a$, ${\bar S}^{\dot a}$.   This establishes that $\Phi(y,\theta)$ is indeed a chiral field generated by the chiral primary operator $A$.  The scaling dimension of the lowest component field $A(x)$ is $\Delta$, as follows from the discussion in Sec.~\ref{sec:rev}.  The $R$-charges of the components are fixed by the $U(1)_R\subset  SU(2,2|1)$ transformation law 
\beq
\Phi'(e^{i\phi/2}X_{\alpha\beta}, e^{5 i \phi/4} \theta_\alpha, e^{2 i \phi} \varphi) = \Phi(X_{\alpha\beta},  \theta_\alpha, \varphi),
\eeq
which implies, together with the definition in Eq.~(\ref{eq:4dchiral}),
\beq
\Phi'(y, e^{3i\phi/4} \theta_a) = e^{i\Delta \phi/2} \Phi(y,\theta).
\eeq
Thus, we recover the proportionality between the $R$-charge of the chiral multiplet and its scaling dimension $\Delta$ at a superconformal fixed point~\cite{Flato:1983te}.  To make contact with the standard normalization of $R$-charge in four-dimensional SCFTs, we may define $R_{4d} = {4\over 3} R$, in which case we have $R_{4d}(\theta_a)=1$ and $R_{4d}(\Phi)= {2\over 3}\Delta$.   The $R$-charges of the component fields are $R_{4d}(A)={2\over 3}\Delta$, $R_{4d}(\psi) = {2\over 3} \Delta-1$, and $R_{4d}(F)={2\over 3}\Delta- 2$.   The property that  $R$-charge is proportional to $\Delta$ in the chiral sector follows from unitarity of the irreducible representations of $SU(2,2|1)$, and in superconformal field theories is reflected in the fact that the $R$-current lives in the same multiplet as the energy-momentum tensor.    In the embedding space approach, this relation is 
a natural consequence of the scaling property $\Phi(\lambda X)=\lambda^{-\Delta}\Phi(X)$ of holomorphic superfields.

\section{Applications to correlators and to the OPE}
\label{sec:apps}

Having established that the embedding space holomorphic operator $\Phi(X)$ gives a covariant representation of the chiral multiplet of  four-dimensional ${\cal N}=1$ SCFTs, we can now determine the consequences of the formalism for SCFT correlation functions.    

A correlator involving arbitrary insertions of holomorphic and anti-holomorphic fields is a function of $SU(2,2|1)$ invariants constructed from the supercoordinates.    It follows from the analysis of $SU(2,2|1)$ Casimir invariants in Ref.~\cite{Jarvis:1978bc} that an (over) complete set of invariants is given by supertraces of products of coordinate supermatrices.    We will sometimes employ the notation 
\beq
\label{eq:sts}
\langle 1 {\bar 2} 3 {\bar 4} \cdots\rangle = \mbox{str} (X_1 \lambda {\bar X}_2   X_3 \lambda {\bar X}_4 \cdots)
\eeq
for such supertraces.

For generic values of the scaling dimensions, correlators that only contain holomorphic field insertions must vanish,
\beq
\langle \Phi_1(X_1)\cdots \Phi_N(X_N)\rangle =0.
\eeq
This follows because tensor products of any number of holomorphic supercoordinates $X_{AB}$ do not contain $SU(2,2|1)$ singlets.   We note however, that for certain cases, non-zero results for purely chiral 2-point and
 3-point~\cite{Molotkov:1975gf,Aneva:1977rj,Conlong:1993eu,Park:1997bq,Osborn:1998qu} and 4-point~\cite{Pickering:1999rk,Dolan:2000uw} functions exist in the literature.   While the formalism presented here does not appear to recover those cases, we note that the chiral 2-point function, which is only non-zero for  dimension $\Delta=3/2$ is purely local (i.e. delta function) and thus not relevant to long distance physics.   The results for the 3-point function in~\cite{Molotkov:1975gf,Aneva:1977rj,Conlong:1993eu,Park:1997bq,Osborn:1998qu} and the 4-point function in~\cite{Pickering:1999rk,Dolan:2000uw} cannot be expressed in a way that simultaneously exhibits manifest holomorphy or superconformal invariance, which is perhaps why they are not present in our formalism\footnote{In addition, the results of~\cite{Molotkov:1975gf,Aneva:1977rj,Conlong:1993eu,Park:1997bq,Osborn:1998qu} on purely chiral 3- and 4-point functions~\cite{Pickering:1999rk,Dolan:2000uw}  apply only to the case in which the dimensions $\Delta_i$ of fields in the correlators obey the constraint $\sum_i\Delta_i =3$.   However, in a unitary CFT, in which dimensions are bounded by $\Delta\geq 1$, these results can only apply to free fields, for which $\Delta=1$.  In that case, these correlators must vanish identically despite being allowed by symmetry.   Indeed, the vanishing of the three-point function has been checked explicitly within the context of a specific model in ref.~\cite{Ferrara:1974fv}.}.

We now consider correlators with both holomorphic and anti-holomorphic insertions.   The simplest object is the two-point function $\langle\Phi_1(X_1) {\bar \Phi}_2({\bar X}_2) \rangle$.   In principle, it is a function of an infinite number of $SU(2,2|1)$ invariants of the form $\langle 1 {\bar 2} 1 {\bar 2} \cdots\rangle$.    However, all these invariants reduce to simple powers of the basic invariant $\langle 1{\bar 2}\rangle= X_1\cdot {\bar X}_2$.   This is most easily seen by applying a superconformal transformation that simultaneously  sends $X_1$ to the origin and ${\bar X}_2$ to conformal infinity.   In that frame, the supertraces reduce to ordinary $SU(2,2)$ invariant traces of products of the matrix $(X_1 {\bar X}_2)_\alpha{}^\beta$, or equivalently, to the $SO(4,2)$ invariant dot product $\eta_{mn} X_1{}^m {\bar X}_2^n$.   Using the scaling property $\Phi_1(\lambda X) = \lambda^{-\Delta_1} \Phi_1(X)$, the only consistent possibility is therefore
\beq
\label{eq:s2pt}
\langle\Phi_1(X_1) {\bar \Phi}_2({\bar X}_2) \rangle = {c_{12}\over \langle 1{\bar 2}\rangle^{\Delta_1}},
\eeq
where, for the same reasons as in the ordinary $SO(4,2)$ case, $c_{12}=0$ unless $\Delta_1=\Delta_2$.    For $X_1$ and  ${\bar X_2}$ on the superlightcone,
\beq
\langle 1{\bar 2}\rangle = -{1\over 2} X_1^+ {\bar X}_2^+ ({\bar y_2} - y_1 + 2 i \theta_1\sigma {\bar \theta}_2)^2,
\eeq
so the four-dimensional two-point function is given by
\beq
\langle \Phi_1(y_1,\theta_1) {\bar \Phi}({\bar y}_2,{\bar\theta}_2)\rangle= {{\hat c}_{12} \over [({\bar y_2} - y_1 + 2 i \theta_1\sigma {\bar \theta}_2)^2]^{\Delta}},
\eeq
where $\Delta=\Delta_1=\Delta_2$.   Expanding both sides in powers of $\theta_1,{\bar\theta}_2$ gives component field two-point functions.   Note that for $\Delta=1$ this is just the superpropagator of a massless chiral field.   For general $\Delta$ this result agrees with previous results~\cite{Molotkov:1975gf,Aneva:1977rj,Park:1997bq,Osborn:1998qu}.

Similar arguments can be applied to the more general case of correlators involving $N$ holomorphic fields $\Phi_i(X_i)$ of dimension $\Delta_i$ and one anti-holomorphic operator ${\bar \Phi}({\bar X})$ of dimension $\Delta$.   The correlator is a function of invariant supertraces of the form, e.g.,
\beq
\label{eq:eg}
\mbox{str}\left (X_1 \lambda {\bar X} X_2 \lambda {\bar X} X_3\lambda {\bar X} \cdots\right).
\eeq
Working in a frame in which $X$ is sent to the origin, one sees that the supertraces reduce to ordinary $SU(2,2)$ traces involving the matrices $(X_i {\bar X})_\alpha{}^\beta$.  Such  traces are expressible in terms of $SO(4,2)$ invariant scalar products of the vectors $X^m_i, {\bar X}^m$ among each other.   Because ${\bar X}^m {\bar X}_m=0$, it follows that the invariants such as those in Eq.~(\ref{eq:eg}) are reducible to products of the basic $SU(2,2|1)$ invariants $X_i \cdot {\bar X}$, not just in the frame with $X$ at the origin, but in an arbitrary frame.    Using the scaling properties  of the holomorphic fields $\Phi_i$, the only possible function of the invariants  $X_i \cdot {\bar X}$ is 
\beq
\label{eq:MHV}
\langle \Phi_1(X_1)\cdots \Phi_N(X_N) {\bar \Phi}({\bar X})\rangle =c_{12\ldots N,{\bar\Phi}} {1\over (X_1\cdot {\bar X})^{\Delta_1}} {1\over (X_2\cdot {\bar X})^{\Delta_2}} \cdots {1\over (X_N\cdot {\bar X})^{\Delta_N}}.
\eeq
Finally, using the scaling behavior of ${\bar \Phi}({\bar X})$, it follows that for non-zero $c_{12\ldots N}$, the  dimensions must obey the constraint $\Delta = \sum_i \Delta_i$, ensuring $R$-charge neutrality of the correlator.  For the special case of $N=2$ holomorphic insertions, Eq.~(\ref{eq:MHV}) is consistent with previous results obtained in refs.~\cite{Molotkov:1975gf,Aneva:1977rj,Park:1997bq,Osborn:1998qu}, but the expression for arbitrary $N$ appears to be new.

The correlator of  Eq.~(\ref{eq:MHV}) has implications for the OPE of chiral operators.   Multiplying the operator product $\Phi_1(X_1\rightarrow X_2) \Phi(X_2)$ by an operator ${\bar\Phi}({\bar X})$ of dimension $\Delta$, and taking vacuum expectation values gives, using Eq.~(\ref{eq:s2pt}) and Eq.~(\ref{eq:MHV}) with $N=2$,
\beq
\Phi_1(X_1\rightarrow X_2) \Phi_2(X_2) \sim  c_{12,{\bar \Phi}} \Phi(X_2) + \cdots,
\eeq
where less singular terms have been omitted.    In this equation, $c_{12,{\bar\Phi}}$ is the numerical coefficient in the 3-point function $\langle \Phi_1(X_1) \Phi_2(X_2) {\bar\Phi}({\bar X})\rangle$, which is non-zero only for $\Delta=\Delta_1+\Delta_2$ (we have chosen an operator basis in which the coefficients of the two-point functions are normalized to unity).   We have recovered the fact that the  OPE of two chiral fields must be non-singular.   In particular, composite operators can be defined multiplicatively, without the need to renormalize UV divergences, and the dimension of the composite operator is the sum of the dimensions of its constituent chiral fields.  On the other hand, the OPE of chiral and anti-chiral fields can have short distance singularities.   If $\Delta_1>\Delta_2$ there is a singularity of the form
\beq
\Phi_1(X_1\rightarrow X_2) {\bar\Phi}_2({\bar X}_2)\sim {\bar c}_{2\Phi,{\bar 1}} (X_1\cdot {\bar X}_2)^{-\Delta_2} \Phi(X_2), 
\eeq
where $\Phi(X)$, if it exists, is an operator of dimension $\Delta_1-\Delta_2$. Meanwhile for $\Delta_2>\Delta_1$, the leading singularity is instead of the form
\beq
\Phi_1(X_1\rightarrow X_2) {\bar\Phi}_2({\bar X}_2)\sim { c}_{1\Phi,{\bar 2}} (X_1\cdot {\bar X}_2)^{-\Delta_1} {\bar\Phi}({\bar X}_2), 
\eeq
where now the dimension of $\Phi(X)$ is $\Delta_2-\Delta_1$.  For $\Delta_1=\Delta_2$, the only consistent possibility is that $\Phi$ is the identity operator.   A more complete description of the subleading terms appearing in the OPE of chiral operators in  ${\cal N}=1$ SCFTs, using four-dimensional language, has been developed in~\cite{Poland:2010wg}.

As a final application, we consider the four-point function $\langle\Phi_1(X_1) \Phi_2(X_2) {\bar \Phi}_3({\bar X}_3) {\bar \Phi}_4({\bar X}_4)\rangle$.  In non-supersymmetric CFTs, the four-point function is not completely fixed by the $SO(4,2)$ symmetry as it can depend on a function of two independent conformally invariant cross ratios.   Taking these to be, e.g., $u={({1\cdot 2}) (3\cdot 4)/ (1\cdot 4) (2\cdot 3)}$ and $v={(1\cdot 3) (2\cdot 4)/ (1\cdot 4) (2\cdot 3)}$,
\bea
\nn
\! \! \! \! \! \! \! \! \!  \langle \Phi_1(X_1) \Phi_2(X_2) \Phi_3(X_3)  \Phi_4(X_4)\rangle &=&  
\left( {2\cdot 3\over (1\cdot 2) (1\cdot 3)}\right)^\frac{\Delta_1}{2}  \left( {1\cdot 3\over (2\cdot 3) (1\cdot 2)}\right)^\frac{\Delta_2}{2}  \\
& & \times  \left( {1\cdot 2\over (1\cdot 3) (2\cdot 3)}\right)^\frac{\Delta_3}{2} \left( {1\cdot 3\over (1\cdot 4) (3\cdot 4)}\right)^\frac{\Delta_4}{2}    f(u,v),
\eea
for some function $f(u,v)$ which in general depends on dynamics and cannot be determined from symmetry alone.

Coming back to the supersymmetric $SU(2,2|1)$ case, the four-point function $\langle\Phi_1(X_1) \Phi_2(X_2) {\bar \Phi}_3({\bar X}_3) {\bar \Phi}_4({\bar X}_4)\rangle$ is in principle a generic function of an infinite number of supertrace invariants such as that in Eq.~(\ref{eq:sts}).   By going to the gauge in which $X_1$ is at the origin and $X_2$ is at infinity, these traces  reduce to ordinary traces of the four matrices $(X_1 {\bar X_3})_\alpha{}^\beta$,  $(X_1 {\bar X_4})_\alpha{}^\beta$,  $(X_2 {\bar X_3})_\alpha{}^\beta$, and  $(X_3 {\bar X_4})_\alpha{}^\beta$.   As before, such traces can be expressed as linear combinations of scalar products of the $SO(4,2)$ vectors $X^m_{i=1,2}$, ${\bar X}^m_{i=3,4}$.    The scalar products involving one $X$ and one ${\bar X}$ are simply the $SU(2,2|1)$ invariant scalar products $X_{i=1,2}\cdot {\bar X}_{j=3,4}$ expressed in our preferred coordinate system.   The traces can also contain factors of the form $(1\cdot 2) ({\bar 3}\cdot{\bar 4})$  (by $R$-invariance, and by the lightcone constraint $X_{i=1,2}^2   = {\bar X}_{i=3,4}^2 =0$, no other combinations are possible).   Using the identity
\beq
4 \, \mbox{tr}\left(X_1 {\bar X}_3 X_2 {\bar X}_4\right) =  (1\cdot {\bar 3}) (2\cdot {\bar 4}) - (1\cdot 2) ({\bar 3}\cdot {\bar 4})  +  (1\cdot {\bar 4}) (2\cdot {\bar 3})
 \eeq
 it follows that, in an arbitrary coordinate system, the complete set of supertrace invariants is generated by the the five independent invariants $\langle i {\bar j}\rangle$ ($i=1,2; j=3,4$)  and $\langle 1 {\bar 3}  2 {\bar 4}\rangle$.   By the usual scaling arguments the 4-point function depends on two independent superconformal invariant cross ratios, which can be chosen as
 \bea
u  &=& {\langle 1 {\bar 3} \rangle \langle 2 {\bar 4}\rangle\over \langle 1{\bar 4}\rangle \langle 2{\bar  3}\rangle}, \\
 v &=& {\langle 1{\bar 3} 2 {\bar 4}\rangle \over \langle 1 {\bar 4} \rangle \langle 2 {\bar 3}\rangle}.
 \eea
Setting $\theta_{i=1,2}={\bar\theta}_{j=3,4}=0$, one sees that $u$ and $v$ become equivalent to the $SO(4,2)$ invariant cross ratios that parametrize the four-point correlator of a generic CFT.   

Given these results, the four-point function for the case of $SU(2,2|1)$ invariance is then of the form
\bea
&& \hspace{-40pt} \langle \Phi_1(X_1) \Phi_2(X_2){\bar \Phi}_3({\bar X}_3) {\bar \Phi}_4({\bar X}_4)\rangle= \nonumber \\
 && \langle 1{\bar 3}\rangle^{{\Delta_2+\Delta_4-3 \Delta_1- 3 \Delta_3}\over 8} \langle 2 {\bar 3}\rangle^{{\Delta_1+\Delta_4-3 \Delta_2- 3 \Delta_3}\over 8} \langle 1{\bar 4}\rangle^{{\Delta_2+\Delta_3-3 \Delta_1- 3 \Delta_4}\over 8} \langle 2{\bar 4}\rangle^{{\Delta_1+\Delta_3-3 \Delta_2- 3 \Delta_4}\over 8}  f(u,v),
\eea
where consistency requires $\Delta_1+\Delta_2=\Delta_3+\Delta_4$.   This expression is not unique as one can always re-define $f(u,v)$ by a power of $u$, which changes the exponents.  The function $f(u,v)$ cannot be fixed by symmetry considerations, although its behavior near certain points is fixed by the OPE.   For example in the limit $X_1\rightarrow X_2$, corresponding to $(u,v)\rightarrow (1,1/2)$, the OPE implies that $f(u=1,v=1/2)=c_{12,{\bar\Phi}} {\bar c}_{34,{\bar\Phi}}$, where $\Phi(X)$ is the operator $\Phi_1(X)\Phi_2(X)$, or
 \beq
\langle \Phi_1(X_1\rightarrow X_2) \Phi_2(X_2){\bar \Phi}_3({\bar X}_3) {\bar \Phi}_4({\bar X}_4)\rangle \rightarrow c_{12,{\bar\Phi}} {\bar c}_{34,{\bar\Phi}} \langle 2{\bar 3}\rangle^{-\Delta_3} \langle 2{\bar 4}\rangle^{-\Delta_4}.
\eeq
The limit $X_1\rightarrow X_3$, with $u,v\rightarrow 0$ implies $f(u,v\rightarrow 0)\sim  c_{1\Phi,{\bar 3}} {\bar c}_{\Phi4,{\bar 2}} u^\frac{-\Delta_1+\Delta_2- 2 \Delta_4}{4}$ for $\Delta_3>\Delta_1$ and $f(u,v\rightarrow0)\sim c_{2\Phi,{\bar 4}} {\bar c}_{3\Phi,{\bar 1}} u^\frac{-\Delta_1-3 \Delta_2+2 \Delta_4}{4}$ for $\Delta_3<\Delta_1$.  Thus,
\beq
\langle \Phi_1(X_1\rightarrow X_3) \Phi_2(X_2){\bar \Phi}_3({\bar X}_3) {\bar \Phi}_4({\bar X}_4)\rangle \rightarrow 
\left\{\begin{array}{cc}
c_{1\Phi,{\bar 3}} {\bar c}_{\Phi4,{\bar 2}} \langle 1{\bar 3}\rangle^{-\Delta_1} \langle 2 {\bar 3}\rangle^{\Delta_1-\Delta_3} \langle 2 {\bar 4}\rangle^{-\Delta_4},
 &  (\Delta_3>\Delta_1)\\
c_{2\Phi,{\bar 4}} {\bar c}_{3\Phi,{\bar 1}}  \langle 1{\bar 3}\rangle^{-\Delta_3} \langle 2 {\bar 4}\rangle^{-\Delta_2} \langle 3 {\bar 4}\rangle^{\Delta_2-\Delta_4}.  &  (\Delta_3<\Delta_1)
\end{array}\right.
\eeq
In the limit $X_1\rightarrow X_4$, corresponding to $u,v\rightarrow\infty$, the asymptotic behavior is $f(u,v\rightarrow\infty)\sim c_{1\Phi,{\bar 4}} {\bar c}_{3\Phi,{\bar 2}} u^\frac{3 \Delta_1+\Delta_2-2\Delta_4}{4}$ for $\Delta_4>\Delta_1$ and $f(u,v\rightarrow\infty)\sim c_{2\Phi,{\bar 3}} {\bar c}_{4\Phi,{\bar 1}} u^\frac{-\Delta_1+\Delta_2 + 2 \Delta_4}{4}$ for $\Delta_4<\Delta_1$, so that
\beq
\langle \Phi_1(X_1\rightarrow X_4) \Phi_2(X_2){\bar \Phi}_3({\bar X}_3) {\bar \Phi}_4({\bar X}_4)\rangle \rightarrow 
\left\{\begin{array}{cc}
 c_{1\Phi,{\bar 4}} {\bar c}_{3\Phi,{\bar 2}}  \langle 1{\bar 4}\rangle^{-\Delta_1} \langle 2 {\bar 3}\rangle^{-\Delta_3} \langle 2 {\bar 4}\rangle^{\Delta_1-\Delta_4},
 &  (\Delta_4>\Delta_1)\\
 c_{2\Phi,{\bar 3}} {\bar c}_{4\Phi,{\bar 1}} \langle 1{\bar 4}\rangle^{-\Delta_4} \langle 4 {\bar 3}\rangle^{\Delta_2-\Delta_3} \langle 2 {\bar 3}\rangle^{-\Delta_2}.  &  (\Delta_4<\Delta_1)
\end{array}\right.
\eeq

\section{Conclusions and Directions for Further Work}
\label{sec:conc}

We have introduced a manifestly covariant formulation of ${\cal N}=1$ SCFTs, based on the lightcone embedding formalism for ordinary CFTs.   The advantages of this framework, as in the $SO(4,2)$ case, is that superconformal transformations act linearly on both coordinates and fields.   This simplifies the construction of invariants of coordinates.   It becomes a simple matter to write down directly the constraints on correlation functions, without the need to solve superconformal Ward identities.

In this paper, we have focused on the relatively simple case of the chiral sector of ${\cal N}=1$ SCFT.   We have shown that holomorphic fields on the projective superlightcone introduced in Sec.~\ref{sec:fields} correspond to chiral ${\cal N}=1$ multiplets of four-dimensional superconformal symmetry.     In Sec.~\ref{sec:apps} we applied the the superlightcone construction to fix the form of correlators with insertions of chiral and anti-chiral operators.    For the 2- and 3-point functions, our results agree with previous results in the literature, while we have obtained new compact expressions for correlators containing an arbitrary number of chiral insertions and one anti-chiral expressions.    For the 4-point function containing two chiral and two anti-chiral fields, we have determined the number of independent superconformal cross ratios that parametrize the dynamics.

We hope that the new simplified kinematics introduced in this paper will lead to new results in superconformal field theory.  Besides the natural extension to non-holomorphic operators and to higher-rank tensor fields, the formalism presented here could be developed in several directions:
\begin{itemize}
\item \emph{Superconformal correlators in curved four-dimensional spaces}:   The $SO(4,2)$ invariant projective lightcone describes not only four-dimensional Minkowski space, but any curved spacetime that is conformally flat.   For instance the intersection of the lightcone with a slice of constant $2 X^6=X^+-X^-$ describes AdS$_4$, with a slice of constant $2 X^4=X^++X^-$ describes dS$_4$, and with a surface of constant $(X^6)^2+(X^0)^2$ describes the space ${\bf S}^1\times {\bf S}^3$.   CFT observables written down in terms of the six-dimensional coordinates $X^m$ (restricted to the lightcone) then automatically describe not just Minkowski space physics but also physics in these other spacetimes (after applying suitable boundary conditions and/or $i\epsilon$ prescription).   Likewise, the superembedding formalism introduced in this paper can be used to study superconformal dynamics in curved spacetime (for some previous work on the case of AdS, which acts as an infrared cutoff~~\cite{Callan:1989em} that preserves supersymmetry, see for instance~\cite{Gripaios:2008rg}).

\item \emph{Extended superconformal symmetry}:  While we have focused on the case of ${\cal N}=1$ superconformal invariance in this paper, it should be possible to extended this framework to study SCFTs with more supersymmetry.   The starting point for this would be spinors of the extended superconformal group $SU(2,2|{\cal N})$,
\beq
V_A =\left(
\begin{array}{c}
V_\alpha\\
\psi^{I=1,\ldots {\cal N}}
\end{array}
\right)
\eeq
and supercoordinates $X_{AB}\sim V_A V_B$ containing the antisymmetric bosonic coordinates $X_{\alpha\beta}$, the symmetric bosonic coordinates $X^{IJ}$, and the fermionic coordinates $\theta^I_\alpha$.   In fact, many of the results presented here already hold if the ${\cal N}=1$ coordinates introduced in Sec.~\ref{sec:6Dsup} are naively replaced with these extended supercoordinates, although the expressions that result upon projection to four dimensions will obviously differ. 

\end{itemize}

We expect to study some of the applications outlined above in future work.

\centerline{\bf Acknowledgements}
We thank J. Maldacena and N. Seiberg for correspondence and comments.   This work is supported in part by DOE grant DE-FG-02-92ER40704, and by an OJI award from the DOE.   WG acknowledges the financial support of the Physics Department at Columbia University, and thanks the theory group at Columbia for its hospitality while this work was being completed. WS would like to thank the Aspen Center for Physics, where part of this work was done.

\begin{appendix}
\section{Spinor conventions}
\label{app:spin}

Fundamental (four-component) spinors of $SU(2,2)$ are written with a lower spinor index,  $V_\alpha$, with $\alpha=1,\ldots,4$.   The conjugate spinor is ${\bar V}_{\dot\alpha}=(V_\alpha)^\dagger$.   Under the (Lorentz) $SL(2,{\bf C})$ subgroup, the fundamental spinor decomposes as
\begin{equation}
V_\alpha =\left(\begin{array}{c}
\psi_a\\
{\bar \chi}^{\dot a}
\end{array}\right),
\end{equation}
where the conventions for two-component spinors $\psi_{a=1,2}$, $\bar{\chi}^{{\dot a}=1,2}$ follow Wess and Bagger~\cite{Wess:1992cp}.   In this basis the $SU(2,2)$ invariant metric $A^{{\dot\alpha}\alpha}$ is defined to be
\begin{equation}
A^{{\dot\alpha}\beta} =\left(
\begin{array}{cc}
0 & {\delta^{\dot a}}_{\dot b}\\
 {\delta_{a}}^{b} & 0
 \end{array}\right).
\end{equation}
This can be used to relate dotted to undotted four-component spinors by ${\bar  V}^\alpha = {\bar V}_{\dot\alpha} A^{{\dot\alpha}\alpha},$ etc.  Finally, the invariant $SU(2,2)$ fully anti-symmetric tensors $ \epsilon_{\alpha\beta\rho\sigma}$,  $\epsilon^{\alpha\beta\rho\sigma}$ are defined by $ \epsilon_{1234}=\epsilon^{1234}=+1$.

The equivalence between $SU(2,2)$ anti-symmetric bi-spinors $X_{\alpha\beta}$,  $X^{\alpha\beta}$ can be made explicit by introducing a basis of  antisymmetric ``sigma matrices" $(\Gamma^{m=+,\mu,-})^{\alpha\beta}$ and ${\tilde \Gamma}^m_{\alpha\beta} = {1\over 2} \epsilon_{\alpha\beta\rho\sigma} \Gamma^{m\rho\sigma}$.   In terms of the usual Dirac matrices of the Lorentz group,
\begin{equation}
{({\gamma^\mu})_\alpha}^\beta = \left(
\begin{array}{cc}
0 & \sigma^\mu\\
{\bar\sigma}^\mu & 0
\end{array}
\right),
\end{equation}
and the charge conjugation matrix
\begin{equation}
C^{\alpha\beta} = \left(\begin{array}{cc}
-\sigma^2 & 0\\
 0 & -\sigma^2
\end{array}
\right),
\end{equation}
these matrices are chosen to be
\begin{eqnarray}
{\Gamma^+}^{\alpha\beta} &=& C^{\alpha\sigma} (1- \gamma_5)_\sigma{}^\beta  
= \left(\begin{array}{cc}
2i\epsilon^{ab} & 0\\
 0 & 0\\
\end{array}
\right),\\
{\Gamma^-}^{\alpha\beta} &=& - C^{\alpha\sigma}(1+\gamma_5)_\sigma{}^\beta = \left(\begin{array}{cc}
0 & 0\\
 0 & 2 i\epsilon_{{\dot a}{\dot b}}\\
\end{array}
\right),\\
{\Gamma^\mu}^{\alpha\beta} &=& iC^{\alpha\sigma}{{\gamma^\mu}_\sigma}^\beta = \left(\begin{array}{cc}
0 & {{\bar\sigma}^{\mu{\dot d} a}} \epsilon_{{\dot d}{\dot b}}\\
- {{\sigma}^\mu}_{d {\dot a}} \epsilon^{{d}{b}} & 0
\end{array}
\right),
\end{eqnarray}
where $\gamma_5 = -i \gamma^0\gamma^1 \gamma^2 \gamma^3=\mbox{diag}(-1,1)$, and
\begin{eqnarray}
{\tilde\Gamma}^+_{\alpha\beta} &=& \left(\begin{array}{cc}
0 & 0\\
 0 & 2 i\epsilon^{{\dot a}{\dot b}}\\
\end{array}
\right),\\
{\tilde\Gamma}^-_{\alpha\beta} &=& \left(\begin{array}{cc}
2 i\epsilon_{ab} & 0\\
 0 &0\\
\end{array}
\right),\\
{\tilde\Gamma}^\mu_{\alpha\beta} &=&
 \left(\begin{array}{cc}
0 & {\sigma}^\mu_ {a{\dot d}} \epsilon^{{\dot d}{\dot b}}\\
- {\bar \sigma}^{\mu  {\dot a} d} \epsilon_{{d}{b}} & 0 
\end{array}
\right).
\end{eqnarray}
These matrices satisfy various useful identities:
\begin{eqnarray}
{\left({\tilde\Gamma}^m \Gamma^n + {\tilde\Gamma}^n \Gamma^m\right)_\alpha}^\beta = -2\eta^{nm}{\delta_\alpha}^\beta,\\
{\left({\Gamma}^m {\tilde \Gamma}^n + {\Gamma}^n {\tilde \Gamma}^m\right)^\alpha}_\beta = -2\eta^{nm}{\delta^\alpha}_\beta,
\end{eqnarray}
\begin{equation}
\Gamma^{m\alpha\beta} {{\tilde \Gamma}^n}_{\alpha\beta} = 4 \eta^{mn},
\end{equation}
\begin{equation}
\Gamma^{m\alpha\beta} {{\tilde \Gamma}_m}{}_{\rho\sigma} = 2\left({\delta_\rho}^\alpha {\delta_\sigma}^\beta - {\delta_\rho}^\beta {\delta_\sigma}^\alpha\right),
\end{equation}
\begin{eqnarray}
\Gamma^{m\alpha\beta} {{\Gamma}_m}^{\rho\sigma} = 2 \epsilon^{\alpha\beta\rho\sigma},\\
{\tilde\Gamma}^m_{\alpha\beta} {\tilde \Gamma}_{m \rho\sigma} = 2 \epsilon_{\alpha\beta\rho\sigma}.
\end{eqnarray}
With these definitions the correspondence between the $SO(4,2)$ vector $X^m$ and the $SU(2,2)$ anti-symmetric bi-spinor $X_{\alpha\beta}$ is 
\begin{eqnarray}
X^{\alpha\beta} = {1\over 2} X_m \Gamma^{m\alpha\beta},& &  X_{\alpha\beta} = \displaystyle{1\over 2} X_m {\tilde\Gamma}^m_{\alpha\beta},
\end{eqnarray}
and 
\begin{equation}
X^m = {1\over 2} X_{\alpha\beta} \Gamma^{m\alpha\beta} =   {1\over 2} X^{\alpha\beta} {\tilde\Gamma}^m_{\alpha\beta}.
\end{equation}
The inner product on vectors is
\begin{equation}
X^m Y_m = X^{\alpha\beta} Y_{\alpha\beta},
\end{equation}
where $SO(4,2)$ vector indices are raised and lowered with the metric $\eta_{mn}$ introduced in the main text and its inverse $\eta^{mn}$.

The matrices $\Gamma^m$ and ${\tilde\Gamma}^m$ can also be used to define a basis of $SU(2,2)$ generators,   
\begin{equation}
{{\Sigma^{mn}}_\alpha}^\beta=-{i\over 4} {\left({\tilde\Gamma}^m \Gamma^n - {\tilde\Gamma}^n \Gamma^m\right)_\alpha}^\beta,
\end{equation}
which satisfies the $SO(4,2)$ algebra
\begin{equation}
[\Sigma^{mn},\Sigma^{pq}] = -i \eta^{nq} \Sigma^{mp} \pm\mbox{perms.}
\end{equation}
The generators ${[{J_\alpha}^\beta]_\rho}^\sigma$ introduced in the main text can be expanded in the $\Sigma^{mn}$ basis and vice-versa 
\begin{eqnarray}
{[{J_\alpha}^\beta]_\rho}^\sigma &=& {1\over 2} {\Sigma_{mn\alpha}}^\beta {{\Sigma^{mn}}_\rho}^\sigma,\\
{{\Sigma^{mn}}_\alpha}^\beta &=&  {{\Sigma^{mn}}_\rho}^\sigma {[{J_\alpha}^\beta]_\sigma}^\rho.
\end{eqnarray}
In particular, an infinitesimal $SU(2,2)$ transformation with parameters ${T_\alpha}^\beta$ in the ${[{J_\alpha}^\beta]_\rho}^\sigma$ basis corresponds to an $SO(4,2)$ transformation generated by $-{i\over 2}\omega_{mn} \Sigma^{mn}$, where 
\begin{equation}
\omega^{mn} =  - {{\Sigma^{mn}}_\beta}^\alpha {T_\alpha}^\beta.
\end{equation}
Finally, the orbital generators ${L_\alpha}^\beta$ and $L_{mn}$ introduced in the main text are related by
\bea
{L_\alpha}^\beta = -  {1\over 2} {{\Sigma^{mn}}_\alpha}^\beta L_{mn}, &  L^{mn} = - {{\Sigma^{mn}}_\alpha}^\beta L_\beta{}^\alpha.
\eea

\section{Differential realization of superconformal generators}
\label{app:diff}

From the explicit transformation rules for the component coordinates $(X_{\alpha\beta},\theta_\alpha,\varphi)$ and their conjugates written in Eqs.~(\ref{eq:t1st})-(\ref{eq:tlast}), we can read off differential operators realizing the algebra on superspace.   The fermionic generators are given by
\begin{equation}
Q_\alpha = X_{\alpha\sigma} {\partial\over\partial\theta_\sigma} + 2\theta_\alpha{\partial\over\partial\varphi} +{\bar\theta}^\sigma{\partial\over\partial {\bar X}^{\alpha\sigma}}-{\bar\varphi} {\partial\over\partial{\bar\theta}^\alpha},
\end{equation}
and
\begin{equation}
{\bar Q}^\alpha = \theta_\beta {\partial\over\partial X_{\alpha\beta}} + \varphi  {\partial\over\partial\theta_\alpha} +{\bar X}^{\alpha\sigma} {\partial\over\partial{\bar\theta}^\sigma}+ 2{\bar\theta}^\alpha {\partial\over\partial{\bar\varphi}}.
\end{equation}
In addition, the bosonic $SU(2,2|1)$ generators are
\begin{equation}
{J_{\alpha}}^\beta ={L_\alpha}^\beta+ \theta_\alpha {\partial\over\partial\theta_\beta} -{\bar\theta}^\alpha{\partial\over\partial{\bar\theta}^\beta} -{1\over 4}{\delta_\alpha}^\beta \left( \theta_\sigma {\partial\over\partial\theta_\sigma} - {\bar\theta}^\sigma{\partial\over\partial{\bar\theta}^\sigma}\right),
\end{equation}
where the ``orbital" part is
\begin{equation}
\label{eq:orbital}
{L_\alpha}^\beta = X_{\alpha\sigma} {\partial\over\partial X_{\beta\sigma}} - {\bar X}^{\alpha\sigma} {\partial\over\partial {\bar X}^{\beta\sigma}} - {1\over 4} {\delta_\alpha}^\beta\left(X_{\rho\sigma} {\partial\over\partial X_{\rho\sigma}} - {\bar X}^{\rho\sigma} {\partial\over\partial {\bar X}^{\rho\sigma}}\right),
\end{equation}
and
\begin{equation}
R= {1\over 4} X_{\alpha\beta} {\partial\over\partial X_{\alpha\beta}} +{5\over 4} \theta_\sigma {\partial\over\partial\theta_\sigma} + {2}\varphi {\partial\over\partial\varphi} -{1\over 4} {\bar X}^{\alpha\beta} {\partial\over\partial {\bar X}^{\alpha\beta}} -{5\over 4}{\bar \theta}^\sigma {\partial\over\partial{\bar \theta}^\sigma} -{2}{\bar\varphi}{\partial\over\partial{\bar\varphi}} 
\end{equation}
generates $U(1)_R$.

Note finally that the differential operators 
\begin{eqnarray}
{\cal D} &=& {1\over 2} X_{\alpha\beta} {\partial\over\partial X_{\alpha\beta}} +\theta_\sigma {\partial\over\partial\theta_\sigma} + \varphi {\partial\over\partial\varphi},\\
{\bar {\cal  D}} &=& {1\over 2} {\bar X}^{\alpha\beta} {\partial\over\partial {\bar X}^{\alpha\beta}} +{\bar \theta}^\sigma {\partial\over\partial{\bar \theta}^\sigma} + {\bar \varphi} {\partial\over\partial{\bar \varphi}},
\end{eqnarray}
commute with the full set of superconformal generators.   They are supersymmetric generalizations of the scaling operator $X^m\partial/\partial X^m$ that plays a role in the projective light cone formulation of $SO(4,2)$ invariant theories.

\end{appendix}

\end{document}